%% file: main.tex
\documentclass[submission,copyright,creativecommons]{eptcs}
\providecommand{\event}{FROM 2025} % Name of the event you are submitting to

\usepackage{iftex}

\ifpdf
  \usepackage{underscore}         % Only needed if you use pdflatex.
  \usepackage[T1]{fontenc}        % Recommended with pdflatex
\else
  \usepackage{breakurl}           % Not needed if you use pdflatex only.
\fi

\usepackage{graphicx}

\usepackage{amsfonts}
\usepackage{amsmath}

\usepackage{array}
\usepackage{booktabs}
\usepackage{tabularx}
\usepackage{caption}
\usepackage{microtype}
\usepackage[frozencache]{minted}
\begin{document}

\newcommand{\smtlib}{SMT-LIB2}
\newcommand{\FP}{\ensuremath{\mathit{FP}}}
\newcommand{\cpp}{C\texttt{++}}

\title{parSAT: Parallel Solving of Floating-Point Satisfiability}
\def\titlerunning{parSAT: Parallel Solving of Floating-Point Satisfiability}
\author{Markus Krahl
\institute{University of Applied Sciences Munich HM\\
Munich, Germany}
\institute{TASKING Germany GmbH\\ Munich, Germany}
\email{markus.krahl@hm.edu\quad markus.krahl@tasking.com}
\and
Matthias Güdemann
\institute{University of Applied Sciences Munich HM\\
Munich, Germany}
\email{matthias.guedemann@hm.edu}
\and
Stefan Wallentowitz
\institute{University of Applied Sciences Munich HM\\
Munich, Germany}
\email{stefan.wallentowitz@hm.edu}
}

\def\authorrunning{M.~Krahl~et~al.}
\maketitle              % typeset the header of the contribution
\input{00_abstract}
\input{01_introduction}
%
\input{02_related}
%
\input{03_theory}
%
\input{04_implementation}
%
\input{05_case_study}
%
\input{06_conclusion}

\bibliographystyle{eptcs}
\bibliography{matthias,paper}

\end{document}

%% file: 00_abstract.tex
\begin{abstract}
    Satisfiability-based verification techniques, leveraging modern
    Boolean satisfiability (SAT) and Satisfiability Modulo Theories (SMT) solvers,
    have demonstrated efficacy in addressing practical problem instances
    within program analysis. 
    However, current SMT solver implementations often
    encounter limitations when addressing non-linear arithmetic problems,
    particularly those involving floating point (FP) operations. This poses a
    significant challenge for safety critical applications,
    where accurate and reliable calculations based on FP numbers and
    elementary mathematical functions are essential.

    This paper shows how an alternative formulation of the satisfiability
    problem for FP calculations allows for exploiting parallelism
    for FP constraint solving. By combining global optimization approaches with
    parallel execution on modern multi-core CPUs, we construct a portfolio-based
    semi-decision procedure specifically tailored to handle FP
    arithmetic. We demonstrate the potential of this approach to complement
    conventional methods through the evaluation of various benchmarks.

\end{abstract}

%% file: 01_introduction.tex
\section{Introduction}
\label{sec:introduction}

Various software-intensive systems utilize Floating Point (FP) computations,
as FP allows for the approximation of real numbers that are
essential in numerous applications modeling physical behavior. When these
systems operate within safety or security critical domains, their correctness
must be guaranteed. The challenge lies in testing and verifying FP computations
due to the vast domain and specific behaviors inherent to FP
arithmetic. However, taking into account recent advancements such as
incorporating IEEE754 compliant FP theory into the \smtlib{} standard, current
state-of-the-art automatic theorem provers offer a viable approach for modeling
and verifying these complex calculations. For solving FP constraints,
contemporary state-of-the-art SMT solvers, e.g., \texttt{bitwuzla}~\cite{niemetzBitwuzla2023},
\texttt{cvc5}~\cite{barbosaCvc5VersatileIndustrialStrength2022}, or
\texttt{Z3}~\cite{demouraZ3EfficientSMT2008},
employ bit-blasting~\cite{brainBuildingBetterBitBlasting2019a}, a technique that
transforms the FP constraints into propositional logic problems. These problems
are then solved by Boolean satisfiability (SAT) solvers, thereby enabling
rigorous, automatic verification of FP computations.

While significant advancements have been made in SMT solver capabilities for
solving equations in FP theory, their capacity for reasoning about FP
constraints, particularly those involving non-linear arithmetic, remains limited
for practical applications. This limitation stems from inherent overhead
associated with converting these computations from FP arithmetic to
propositional logic, often requiring word-~\cite{niemetzBitwuzla2023}
or bit-blasting operations specific to the FP-focused solver
frameworks. Furthermore, this encoding process fails to exploit advancements in
hardware such as dedicated FP units within multicore CPUs or GPUs that could
significantly accelerate computation.

In this paper, we present parSAT, an integrated tool that performs a
portfolio-based semi-decision procedure for SMT equations in FP theory.
It extends the formulation of solving FP constraints based on global
optimization (GO), originally introduced and refined in
\cite{fuXSatFastFloatingPoint,benkhadraGoSATFloatingpointSatisfiability2017}. We
specifically address limitations by incorporating a more comprehensive support
for \smtlib{} constraints and enabling a portfolio-based minimization of the
objective function using a multicore implementation to decide satisfiability.
Furthermore, we demonstrate the potential of combining parSAT with an SMT solver
to systematically leverage the advantages of both approaches.

The paper is structured as follows. Section~\ref{sec:related-work} discusses
related work, and section~\ref{sec:theory} provides the theoretical foundations
for converting SMT equations in FP theory.  The core principles and
implementation details of parSAT are described in
section~\ref{sec:implementation}. In section~\ref{sec:evaluation}, we evaluate
parSAT in comparison with other SMT solvers. Section~\ref{sec:conclusion}
concludes the paper and discusses the outlook and potential further work.

%%% Local Variables:
%%% mode: latex
%%% TeX-master: "main"
%%% End:

%% file: 02_related.tex
\section{Related Work}
\label{sec:related-work}

The approach to formulate FP constraints in the form of a GO
problem was first introduced in XSat~\cite{fuXSatFastFloatingPoint}. It solves
SMT equations in pure FP theory by converting the given constraints into an
equisatisfiable mathematical optimization problem which has a global minimum of
value 0 if and only if the original problem is satisfiable. In detail, it
transforms a quantifier-free SMT equation $F(\overrightarrow{x})$, where
$\overrightarrow{x} \in FP^{n}$ corresponds to an arbitrary assignment of the
variables in the equation, into an objective function $G(\overrightarrow{x})$.
$G(\overrightarrow{x})$ is minimized by applying GO to find
an input vector $\overrightarrow{z}$ for which $G(\overrightarrow{z}) = 0$. In
case a $\overrightarrow{z}$ was found, $\overrightarrow{z}$ would correspond to
a valid assignment $\alpha$ for which $F(\alpha) = \textit{SAT}$, therefore
$F(\overrightarrow{x})$ would be satisfiable. In case only a non-zero global
minimum is found, the SMT equation $F(\overrightarrow{x})$ is considered to be
unsatisfiable.  XSat takes an SMT equation in the \smtlib{} format as input and
generates the optimization function as C-code which is compiled and loaded as
Python-Module that applies the GO algorithm
\texttt{Basin Hopping}~\cite{walesGlobalOptimizationBasinHopping1997}
(\texttt{BH}) from the SciPy-Package~\cite{2020SciPy-NMeth} to find its global minimum.

goSAT~\cite{benkhadraGoSATFloatingpointSatisfiability2017} is based on the ideas
of XSat as it similarly transforms an SMT equation in pure FP theory from a given
input file in \smtlib{} format to an optimization function and attempts to find
its global minimum.  It also supports a code generation mode similar to XSat
where the optimization function is emitted as C-code. However, the main
aspect of goSAT is the Just-in-Time (JIT) compilation of the generated objective
function and the immediate search for its global minimum. Additionally, it
offers the possibility to choose from several different GO routines provided
through the NLopt~\cite{johnsonNLoptNonlinearOptimization2021} library
whereas XSat only involves the \texttt{BH} algorithm.

Our parSAT approach builds on top of XSat and goSAT. We included the three best performing GO
algorithms based on average runtime and found \textit{SAT}
solutions used in goSAT and XSat, namely \texttt{BH} with the \texttt{Powell} method~\cite{powell-method}
as local minimizer, \texttt{Controlled Random Search 2 with local mutation}
~\cite{priceGlobalOptimizationControlled1983,kaeloVariantsControlledRandom2006} (\texttt{CRS2}) and
\texttt{Improved Stochastic Ranking Evolution Strategy}~\cite{isres-paper} (\texttt{ISRES})
for use in minimization by reimplementing each approach in \cpp. parSAT supports
a freely configurable parallel execution of these algorithms in a
portfolio-based approach. Similar to XSat and goSAT, parSAT first converts the
given SMT equation into an optimization function, but it natively compiles this
optimization function into a shared library that is directly loaded into parSAT.
Additionally, parSAT provides a more complete support of the \smtlib{} standard,
e.g., the \textit{ite} function for arguments with FP type.

Most other SMT solvers apply the
DPLL(T)~\cite[p.~66]{kroeningDecisionProcedures2016} framework which is used to
decide combinations of different theories based on structural SAT solving. In
that case, an SMT solver consists of multiple theory solvers and an
orchestrating SAT solver. The SMT solvers
\texttt{bitwuzla}~\cite{niemetzBitwuzla2023} and
\texttt{cvc5}~\cite{barbosaCvc5VersatileIndustrialStrength2022} use an implementation of
FP constraints based on the SymFPU~\cite{brainBuildingBetterBitBlasting2019}
library. This allows them to apply word-blasting, i.e., an eager conversion of
FP into the bitvector theory and then using bit-blasting to generate a purely
propositional bit-level problem for a SAT solver. It can be beneficial to not
only have conjunctions for the theory solver, but to use more abstract reasoning
on the mathematical level in order to avoid searching through large state spaces
that are impossible according to the theory. This is for example realized in the
natural domain solving procedure implemented in
\texttt{MathSAT}~\cite{hallerDecidingFloatingpointLogic2012} or based on incremental
approximations and logic programming as in
Colibri~\cite{zitounEfficientConstraintBased2020}.
These approaches to solving FP constraints are powerful and guaranteed to be
complete. The downside is that due to the complexity of the underlying
propositional bit-vector formulas, finding a solution in practical problems is
often difficult. They also do not exploit modern multi-core CPUs or GPUs with
dedicated FP units to accelerate FP instructions which
is possible using the parSAT approach.

Previous research has explored running SMT solvers in a portfolio setting,
where multiple configurations of a single solver or different solvers are
executed in parallel to solve a given equation, such as~\cite{concurrent-portfolio-approach},
\cite{kovacs-portfolio-sat-smt}, and \cite{weber-tjarkpar-2023}.
Our approach similarly adopts a portfolio strategy; however,
instead of using multiple SMT solvers, we concurrently execute different
GO methods to locate a zero-valued global minimum of the optimization function.
Since all instances operate within the same optimization framework,
previously evaluated points could be shared more efficiently among
them compared to running separate SMT solvers.

%%% Local Variables:
%%% mode: latex
%%% TeX-master: "main"
%%% End:

%% file: 03_theory.tex
\section{Theoretical Background}
\label{sec:theory}

In this chapter, we provide the theoretical foundation of parSAT. Let \FP{} be
the set of IEEE754 \texttt{double} precision FP numbers. Note that
this set includes the numbers that can be represented precisely by single
precision numbers, i.e., of type \texttt{float}.

In general, a quantifier-free SMT equation $F(\overrightarrow{x})$ with
$\overrightarrow{x} \in \FP^{n}$ is transformed into a mathematical objective
function $G(\overrightarrow{x})$ in such a way that computing
$G(\overrightarrow{x})$ with a given input vector $\overrightarrow{a}$ either
returns $0$ if $\overrightarrow{a}$ corresponds to a \emph{satisfiable}
assignment $\alpha$ for $F(\overrightarrow{x})$ or a positive distance value
that indicates how close $\overrightarrow{a}$ is to a global minimum at zero,
i.e., a satisfiable assignment.

To ensure the equivalence between the optimization function
$G(\overrightarrow{x})$ and the initial SMT FP formula $F(\overrightarrow{x})$,
the following requirements, originally stated in XSat, must hold: \\

\noindent\textbf{R(1)}: $\forall \overrightarrow{x} \in \FP^{n} \, \to \, G(\overrightarrow{x}) \geq 0$\\
\textbf{R(2)}: $\forall \overrightarrow{x} \in \FP^{n} \land G(\overrightarrow{x}) = 0 \, \to \, \overrightarrow{x} \models F$\\
\textbf{R(3)}: $\forall \overrightarrow{x} \in \FP^{n} \land \overrightarrow{x} \models F \, \to \, G(\overrightarrow{x}) = 0$\\

\noindent\textbf{R(1)} states that the objective form is non-negative. \textbf{R(2)}
states that if the objective function has a value of $0$ then the corresponding
valuation of the free variables in the constraint problem is a satisfying
assignment. Finally \textbf{R(3)} states that every satisfying assignment to the
constraint problem corresponds to a root of the objective function.

parSAT supports a given SMT formula $F(\overrightarrow{x})$ in the language
$L_{\FP-SMT}$ representing quantifier-free FP constraints. The
language $L_{\FP-SMT}$ supported by parSAT is a strict superset of the language
supported by goSAT and XSat, its syntax is defined as:\\
\\

\noindent \textbf{\textit{Boolean constraints}} \quad $\pi := \neg\pi' \, | \, \pi_1 \land \pi_2 \, | \, \pi_1 \lor \pi_2 \, | \, e_1 \bowtie e_2$\\
\textbf{\textit{FP expressions}} \,\,\qquad\quad $e := c \, | \, v \, | \,  e_1 \otimes e_2 \, | \,fun(e_1,...,e_n) \, | \, ite(\pi, e_1, e_2)$\\

\noindent where $\bowtie \in \{  <, \leq, >, \geq, ==, \neq \}$,$\otimes \in \{  +, -, *, / \}$, \textit{c} represents a FP constant,
$v$ is a FP variable, \textit{fun} is a user-defined, interpreted FP function, and \textit{ite} corresponds to the \textit{if-then-else}-function
defined in the \smtlib{} core theory. It returns the FP expression $e_1$ if the Boolean argument $\pi$  is true;
otherwise, it returns the FP expression $e_2$.

Similar to goSAT, we define $F_{CD}(\overrightarrow{x})$ as the conversion of
$F(\overrightarrow{x})$ by removing the negation through the application of
De-Morgan's law and transforming it into \textit{conjunctive normal form}. But
in contrast to the approach described for goSAT, we denote each
operand in $\bowtie$ with an additional subscript $n$ to indicate whether the
initial operand is negated due to the application of De-Morgan's law:

\begin{equation}
    \label{eq:f-cnf}
    F_{CD}(\overrightarrow{x}) = \bigwedge_{i \in I}\bigvee_{j \in J}e_{i,j}\bowtie_{i,j,n}e'_{i,j}
\end{equation}

From $F_{CD}(\overrightarrow{x})$ we deduce the optimization function $G(\overrightarrow{x})$ as follows:

\begin{equation}
    \label{eq:f-opt}
    G(\overrightarrow{x}) = \sum_{i \in I}\prod_{j \in J}d(\bowtie_{i,j,n}, e_{i,j}, e'_{i,j})
\end{equation}

\noindent where $d(\bowtie_{i,j,n}, e_{i,j}, e'_{i,j})$ translates the boolean value of the comparison operators in
$\bowtie$ to an FP value that is equal to or greater than zero.

Here, the previously introduced subscript $n$ for an intended negation of the comparison operator needs to
be considered. For instance, for the real numbers $r_a,r_b$ the statement $\neg (r_a < r_b)$
could be transformed to $(r_a \geq r_b)$ to eliminate the negation.
However, this transformation would not be valid for all FP values, specifically the \textit{NaN} value.
In case for the FP numbers $f_a, f_b$ where $f_b = \textit{NaN}$, the first statement $\neg (f_a < f_b)$ would be
true, whereas the second statement $(f_a \geq f_b)$ would be false.
In order to address this special case in FP arithmetic, we insert an additional check
for $\textit{NaN}$ to $d(\bowtie_{i,j,n}, e_{i,j}, e'_{i,j})$ when
the corresponding operator in $\bowtie$ was later negated due to the application of the De-Morgan's law,
indicated by subscript $n = 1$.

This results in the following definitions to encode boolean constraints involving
FP expressions,

\begin{align}
    d(==_0, e_1, e_2) & = d(\neq_1, e_1, e_2) = \theta(e_1,
                                                  e_2)\label{eq:d-eq} \\
    d(\neq_0, e_1, e_2) & = d(==_1, e_1, e_2) = e_1 \neq e_2 \,\, ? \,\, 0 \, :
                                                 \, 1    \label{eq:d-neq} \\
    d(<_0, e_1, e_2) &= e_1 < e_2 \,\, ? \,\, 0 \, : \, \theta(e_1, e_2) +
                       1 \label{eq:d-lt}\\
    d(<_1, e_1, e_2) &= isnan(e_1)  \lor isnan(e_2)  \,\, ? \,\, 0  \, :  \,
  d(\geq_0, e_1, e_2)    \label{eq:d-not-lt}\\
    d(\leq_0, e_1, e_2) &= e_1 \leq e_2 \,\, ? \,\, 0 \, : \, \theta(e_1,
  e_2)    \label{eq:d-leq}\\
    d(\leq_1, e_1, e_2) &= isnan(e_1)  \lor isnan(e_2)  \,\, ? \,\, 0  \, :  \,
  d(>_0, e_1, e_2)    \label{eq:d-not-leq}\\
    d(>_0, e_1, e_2) &= e_1 > e_2 \,\, ? \,\, 0 \, : \, \theta(e_1, e_2) +
  1    \label{eq:d-gt}\\
    d(>_1, e_1, e_2) &= isnan(e_1)  \lor isnan(e_2)  \,\, ? \,\, 0  \, :  \,
  d(\leq_0, e_1, e_2)    \label{eq:d-not-gt}\\
    d(\geq_0, e_1, e_2) &= e_1 \geq e_2 \,\, ? \,\, 0 \, : \, \theta(e_1, e_2)
    \label{eq:d-geq}\\
    d(\geq_1, e_1, e_2) &= isnan(e_1)  \lor isnan(e_2)  \,\, ? \,\, 0  \, :  \, d(<_0, e_1, e_2)    \label{eq:d-not-geq}
\end{align}

\noindent where $\theta(f_1, f_2)$ computes the distance value between the FP
numbers $f_1$ and $f_2$. Similarly to goSAT, we calculate $\theta(f_1, f_2)$
where $f_1, f_2 \in \FP \setminus \{\textit{NaN}\}$ with the difference in the
bit representations of $f_1$ and $f_2$.  To adhere to the IEEE754 standard, that
requires that the equality comparison of two \textit{NaN} values is always
false, \textit{NaN} values need to be treated differently as the difference in
the bit representation of two \textit{NaN} values might be zero. Therefore, the
following definition is used in parSAT:
\begin{equation}
    \label{eq:theta}
    \theta(e_1, e_2) = isnan(e_1)  \lor isnan(e_2)  \,\, ? \,\, 1  \, :  \, (e_1 == e_2 \,\,  ? \,\,  0 : \, (|bits(e_1) - bits(e_2)|))
\end{equation}
\noindent Accordingly, the following properties hold:
\begin{align}
  \forall f_1, f_2 \in \FP: \quad & \theta(f_1, f_2) \geq 0    \label{eq:theta-prop1} \\
  \forall f_1, f_2 \in \FP: \quad & \theta(f_1, f_2) = 0 \implies \, f_1 =
                                   f_2    \label{eq:theta-prop2} \\
  \forall f_1, f_2 \in \FP: \quad & \theta(f_1, f_2) = \theta(f_2,
                                   f_1)    \label{eq:theta-prop3}\\
  \forall f_{1},f_{2}\in \FP: \quad & isnan(f_{1})  \lor isnan(f_{2}) \implies \, \theta(f_1, f_2) > 0    \label{eq:theta-prop4}
\end{align}

\noindent Considering equations (2) to (17), we can derive that the previously stated requirements \textbf{R(1)}, \textbf{R(2)},
and \textbf{R(3)} hold for $G(\overrightarrow{x})$.
Because of the distinct processing of negated comparison operators, parSAT also enables the generation of
optimization functions that correctly return zero for an assignment $\alpha$ containing \textit{NaN} values,
which satisfies the initial SMT FP constraints.
Still, GO algorithms might be developed based on mathematical reasoning and therefore might be limited
to calculate global minima that do not contain infinite FP numbers, such as \textit{NaN} or \textit{Infinity}
even though these might present a valid solution for the given SMT equation.

FP expressions are directly encoded by performing the operations defined by $\otimes$
using the corresponding \texttt{double} or \texttt{float} types in C.
In this way, the potential non-linear characteristics of the original SMT
equation are preserved when translating it into the generated GO function.

Generally, parSAT allows the integration of any derivative-free GO algorithm.
Due to the various conditionals in the previously presented equations that construct
the generated optimization function, it is very likely that the final optimization function is
not smooth and therefore not differentiable.
All GO procedures currently incorporated in parSAT are
employing a stochastic driven strategy for locating a global minimum.
Therefore, parSAT fulfills the soundness property only for SMT equations where one of its GO algorithm finds a zero valued
minimum $\overrightarrow{z}$, since $\overrightarrow{z}$ necessarily represents
a valid solution to the initial SMT formula $F(\overrightarrow{x})$.
However, parSAT is incomplete and unsound for SMT equations where it cannot find a satisfiable assignment
due to the non-deterministic behavior of the employed GO algorithms.
Because of their stochastic nature, the GO algorithms employed by parSAT may not find a zero-valued minimum with respect to the given time
and evaluation limits, even though the optimization function has a global
minimum of zero, i.e., the given SMT formula is satisfiable.
Therefore, when parSAT cannot find a solution for the generated optimization functions,
it is unable to reason if the given SMT problem is either satisfiable or unsatisfiable and emits \textit{UNKNOWN}.
Accordingly, parSAT may only find a satisfiable assignment for a given SMT equation but cannot prove unsatisfiability.

%%% Local Variables:
%%% mode: latex
%%% TeX-master: "main"
%%% End:

%% file: 04_implementation.tex
\section{Implementation Considerations}
\label{sec:implementation}

\begin{figure}[ht]
  \centering
  \includegraphics[width=0.9\textwidth]{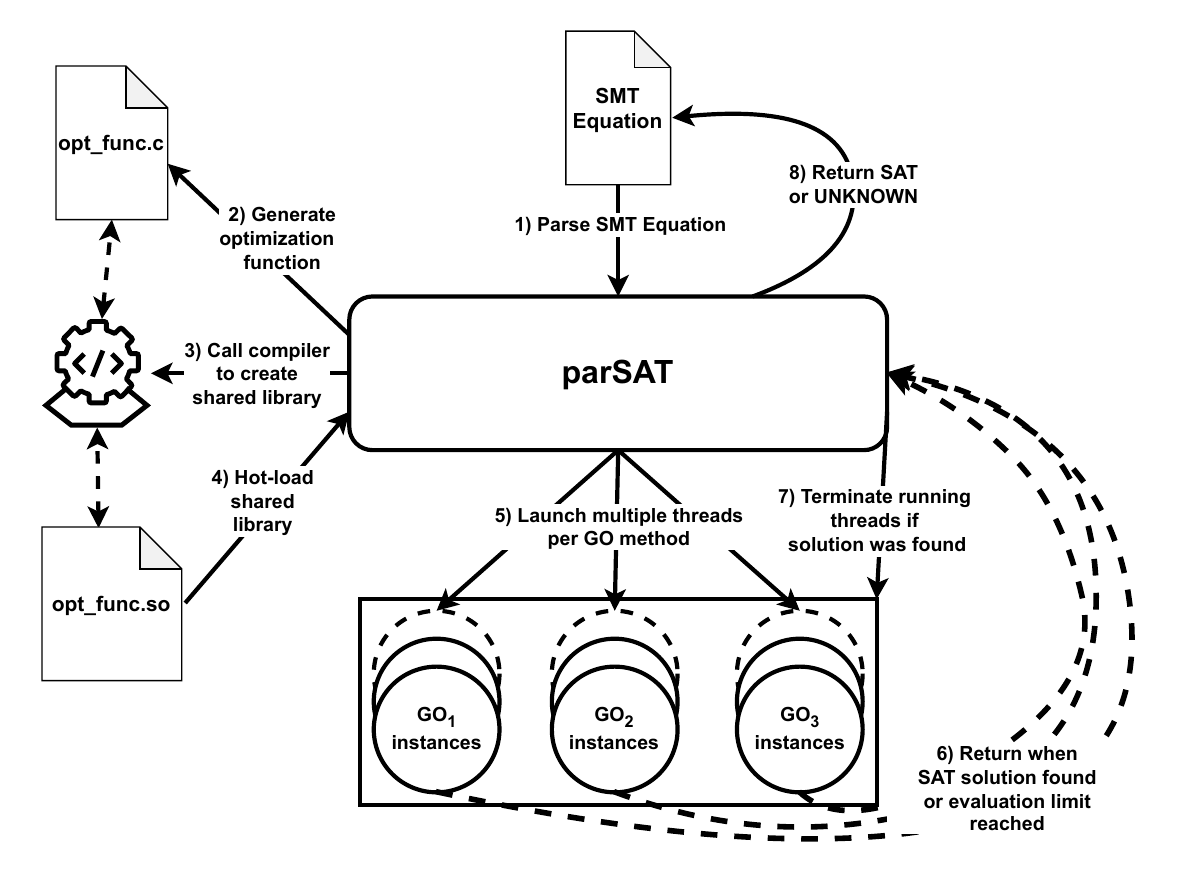}
  \caption{Overview of parSAT's execution behavior}
  \label{fig:parsat-overview}
\end{figure}

In the following, we present the core implementation decisions of parSAT.
First, we elaborate on the method used to parse a given SMT equation and describe how
the resulting optimization function is incorporated into parSAT's subsequent execution process.
Second, we examine the design principles that facilitate the parallel execution of multiple
GO algorithms, which enable a parallel, distributed approach to solving the initial FP equation.
We analyze the current set of GO algorithms integrated into parSAT,
including their parameter settings and the potential number of instances
that may be created for each algorithm.
Finally, we provide a brief example how a generated GO function by parSAT would compare
to the initial FP satisfiability problem.
parSAT and all of its integrated GO algorithms are written in \cpp.
Figure~\ref{fig:parsat-overview} provides an overview of the operation process of parSAT.
The details concerning each execution step are described in the following
subsections.

\subsection{Optimization Function Generation and Compilation}
parSAT accepts an SMT file compliant to the \smtlib{} standard as
input. Currently only the quantifier-free FP theory is supported. The rounding
mode for the FP operations is set to \emph{round nearest ties to even} (RNE).
Support for the other rounding modes can be added by inserting calls to the
corresponding functions defined in the \texttt{fenv}-header of the C standard
library when generating the C code for the optimization function.

We use the parser in \texttt{libz3} from the \texttt{Z3} SMT solver to parse
each \textit{assert} statement of the SMT file into a parse tree with all
involved FP variables, constants and operations affecting a particular
\textit{assert} statement. The root node represents the constraint whereas each
node corresponds to either a Boolean or FP operation and each leaf represents
either a FP variable or constant. We also apply the \texttt{simplify} method of
\texttt{libz3} to optimize each assertion tree, e.g., to eliminate unused
symbols or redundant definitions.

Afterwards, we recursively iterate through each assertion tree to generate the optimization function as C code.
First, the C variables for the leaves, i.e., FP constants or variables, are created.
Second, starting from the closest leaf-nodes to the assertion
top-node further variables are defined that connect the variable names of the children with the
corresponding operator until finally the variable of the assertion node is constructed.
Accordingly, each assertion variable defines an optimization function $G_a(\overrightarrow{x})$
in semantic equivalence to the equation (2) to (17).
Finally, the overall result is calculated by putting all assertions in conjunction.
Because of equation (2) the final result of the optimization function is the sum of all assertion
optimization functions such that $G(\overrightarrow{x}) = \sum_{a=1}^{I}v_a$ where each assertion
variable $v_a$ represents $G_a(\overrightarrow{x})$ and $I$ equals to the amount of assertions
in the given SMT file.

To avoid rounding errors introduced during the transformation process we assign single precision FP
symbols to variables of type \texttt{float} and double precision FP numerals to \texttt{double} typed
C variables. Other sorts of FP types, such as half precision (16-bit) or quadruple precision (128-bit)
are currently not supported. However, the input vector given to the optimization function and iteratively modified
by the GO algorithms is typed as \texttt{double} since FP values in double precision are capable of
accurately representing single precision numbers.
Because of this, the employed GO algorithms could actually
calculate global minima that contain $\pm \infty$ values in one of their coordinates if this particular
dimension corresponds to a single precision FP number in the initial SMT equation.

The fully generated optimization function is written into a source file and compiled as shared library.
Besides the necessary flags to compile the generated GO function as shared library,
the default settings of the compiler (here \texttt{gcc}) are utilized.
Therefore, FP operations shall be performed according to the IEEE754 standard.
We use the \texttt{system} function call of \cpp{} to invoke the compiler and, after successful compilation,
the \texttt{dlopen} and \texttt{dlsym} functions to immediately load the compiled optimization function
in the execution context of parSAT.
Subsequently, a reference to the optimization function is passed to each GO instance.
Since the optimization function is a pure function, i.e., it has no side effects, it can be safely
called by multiple threads running in parallel.

\subsection{Parallel GO Procedure}
Generally, parSAT launches a freely configurable number of GO instances (with respect to the resources
of the available runtime environment) that concurrently search
for a global minimum of the provided optimization function $G(\overrightarrow{x})$.
This may also include multiple instances of the same GO method.

parSAT terminates when the first thread finds a minimum equal to zero, signaling
that a satisfiable assignment for the input formula has been found, or when each
GO instance has reached its evaluation limit, indicating that it is \textit{UNKNOWN} whether
the given SMT equation is satisfiable. The evaluation limit sets a maximum number for how many
times each GO method may call the previously generated GO function.

We reimplemented \texttt{CRS2} and \texttt{ISRES} with the available code of NLopt
and the algorithm descriptions in the original papers as reference but adopted the approaches
to leverage the unique properties of parSAT's generated optimization functions.
For \texttt{BH} with the \texttt{Powell} method as local minimizer,
we manually converted the Python code from SciPy into \cpp.
We slightly modified the \texttt{Powell}'s minimization method to avoid introducing non-finite FP values,
such as $\pm \infty$ or \textit{NaN}, during the minimization process, possibly induced because of large deltas between subsequent minima.
As source for returning random double values required by the employed GO algorithms,
we implemented a pseudo random number generator based on the \texttt{xoshiro256+}~\cite{xoshiro256} algorithm.

Currently, parSAT does not distribute the work for a single GO method into multiple threads,
but executes each GO process together with the optimization function in a distinct thread.
Due to the large search space of one- to multidimensional GO problems in the FP domain,
we did not implement a synchronization mechanism to share the already evaluated coordinates between
the GO threads.
However, at the beginning of the solving process, parSAT generates for each GO instance
a randomized input vector to ideally have a broad distributed
starting set between all concurrent running GO processes.
We included stochastic GO methods into parSAT where running the same
GO algorithm in multiple distributed instances would randomly evaluate different regions
of the search space.
This approach shall also avoid that multiple GO instances are being trapped in the same local minimum.

\subsection{Exemplary Execution of parSAT}
To demonstrate the execution process of parSAT, we employ the following quadratic
equation \\

\noindent$ t(x) = -1 \dot (x + 2)^2 - 2$\\

\noindent for which the following exemplary property shall hold:

\noindent\textbf{ExP(1)}: $ x \in \FP \land t(x) \geq -2 $.\\

\noindent When \textbf{ExP(1)} and $t(x)$ are encoded as SMT equation in FP theory
(as presented in Listing~\ref{lst:smt-equation}) and submitted to parSAT, it generates a
GO function where its integrated GO algorithms would attempt to find
a global minimum of zero.

\definecolor{verylightgray}{rgb}{.95,.95,.95}

\begin{listing}[H]
  \small
\begin{minted}[linenos,bgcolor=verylightgray]{lisp}
(set-logic QF_FP)
;; set rounding mode to round to nearest, ties to even
(define-fun rm()    RoundingMode RNE)
;; symbolic variable for input parameter x
(declare-fun x()    (_ FloatingPoint 8 24))
;; a:= -1.0 ; x_s := 2.0; y_s := -2.0; max_y := -2.0
(define-fun a()     (_ FloatingPoint 8 24) ((_ to_fp 8 24) #xbf800000))
(define-fun x_s()   (_ FloatingPoint 8 24) ((_ to_fp 8 24) #x40000000))
(define-fun y_s()   (_ FloatingPoint 8 24) ((_ to_fp 8 24) #xc0000000))
(define-fun max_y() (_ FloatingPoint 8 24) ((_ to_fp 8 24) #xc0000000))
(define-fun x2_1()  (_ FloatingPoint 8 24) (fp.add rm x x_s))
(define-fun x2_2()  (_ FloatingPoint 8 24) (fp.mul rm x2_1 x2_1))
(define-fun x2_3()  (_ FloatingPoint 8 24) (fp.mul rm a x2_2))
(define-fun x2_4()  (_ FloatingPoint 8 24) (fp.add rm x2_3 y_s))
;; constrain the possible solution to satisfy the property
(assert (fp.geq x2_4 max_y))
;; check if the problem has a solution
(check-sat)
\end{minted}
  \normalsize
  \caption{\label{lst:smt-equation} SMT equation of \textbf{ExP(1)} and $t(x)$}
\end{listing}

\begin{figure}[h!]
  \centering
  \includegraphics[width=0.8\textwidth]{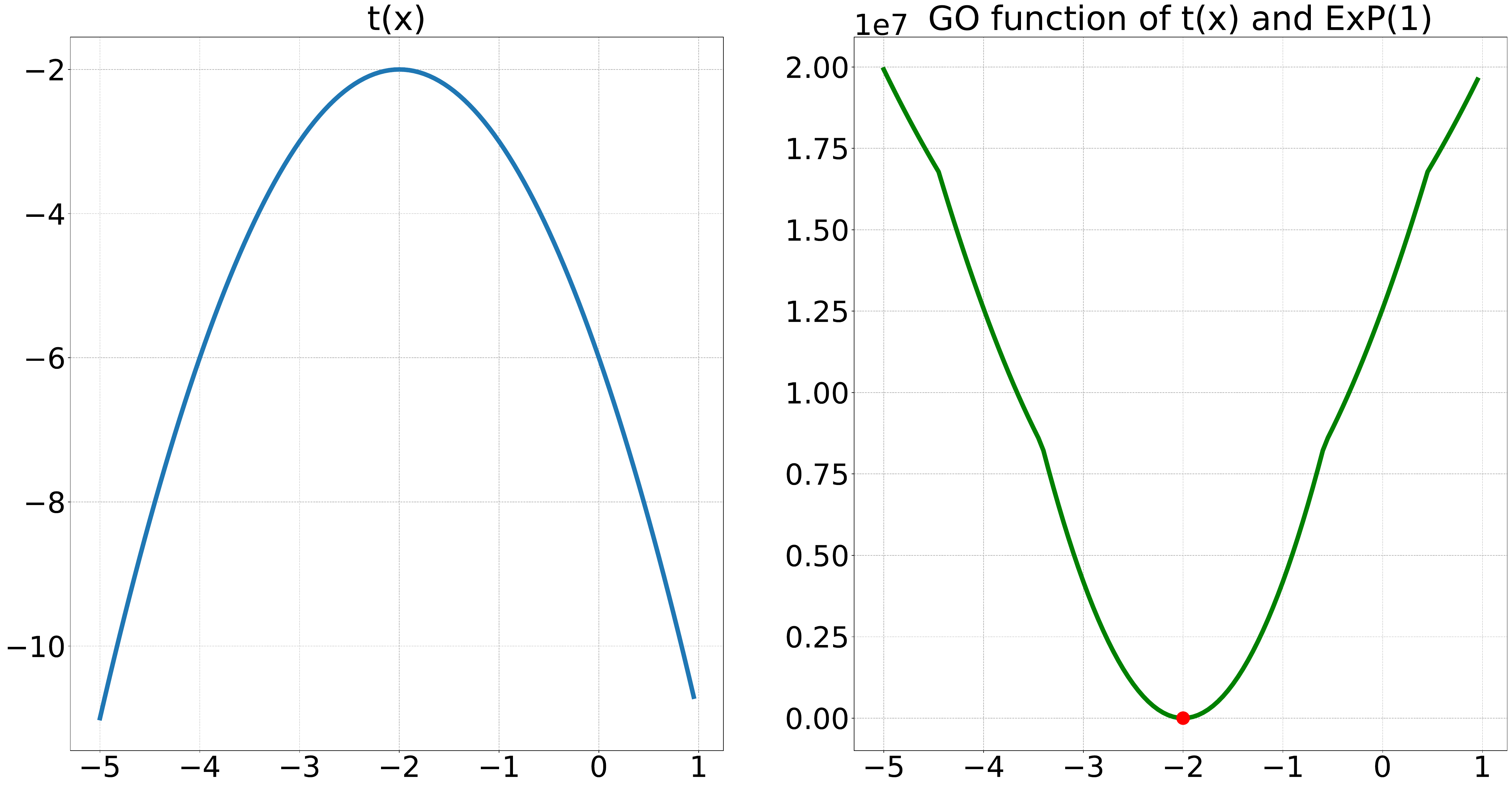}
  \caption{Function plots of t(x) and the generated GO function}
  \label{fig:example-go-plot}
\end{figure}

Figure~\ref{fig:example-go-plot} illustrates the function $t(x)$ (left panel) and
the plot of the corresponding GO function (right panel) generated by parSAT based on Listing~\ref{lst:smt-equation}.
The construction of the GO function follows the set of equations (2-17),
with equation (11) specifically employed to encode the \texttt{assert} statement involving
a greater-than-or-equal-to relational operator.
As observed, the GO function exhibits a local minimum of $0.0$
at $x = -2.0$, indicated by the red marker.
At this point, $t(-2.0) = -2.0$, thereby satisfying the constraint
\textbf{ExP(1)}: $t(-2.0) \geq -2.0$.
Therefore, $x = -2.0$ constitutes a satisfiable assignment for the SMT equation
in Listing~\ref{lst:smt-equation}.

%%% Local Variables:
%%% mode: latex
%%% TeX-master: "main"
%%% End:

%% file: 05_case_study.tex
\section{Evaluation}
\label{sec:evaluation}

We evaluated the efficiency and effectiveness of parSAT on the complete Griggio
benchmark set, which is a subset of the quantifier-free FP benchmarks referenced
by the \smtlib{} initiative~\cite{preiner-2024-11061097}.
The Griggio benchmark containing 214 SMT instances
was added to the benchmark set of the \smtlib{} initiative
after the FP theory was supported by the \smtlib{} standard in 2015.
Since then, this benchmark was often used to evaluate the effectiveness and efficiency
of SMT solvers on handling FP constraints such as in~\cite{fuXSatFastFloatingPoint}
and ~\cite{benkhadraGoSATFloatingpointSatisfiability2017}.
Additionally, we analyzed the performance of parSAT on a second benchmark, the
2019-Guedemann benchmark set, which is based on verifying calculations to
approximate the natural logarithm and the exponential functions, and which became a
subset of the quantifier-free FP and linear real arithmetic benchmarks maintained by
\smtlib.

\subsection{Overview}
\label{sec:overview}

Firstly, we show the results of analyzing various parSAT variants by employing
different parallelization strategies for the included GO algorithms. Secondly,
we compare the best observed version of parSAT against current state-of-the-art
SMT solvers, including \texttt{bitwuzla} (0.7.0), \texttt{cvc5} (1.2.1),
\texttt{Z3} (4.13.4), and \texttt{MathSAT} (5.6.11). Each solver was executed
with its default parameters on the Griggio benchmark.
Secondly, we examine the results of running the same set of SMT solvers with their
default configuration and parSAT on the 2019-Guedemann benchmark.
Finally, we propose a combined approach that integrates parSAT with an SMT solver,
aiming to harness the strengths of both methods to establish a fast and reliable solving process.

In all experiments, the primary evaluation criteria were the efficiency and solution quality of the approach being investigated,
i.e., a particular configuration of parSAT or an SMT solver.
The solution quality was assessed based on the number of \textit{SAT} instances identified within the benchmark;
a higher count of discovered \textit{SAT} instances indicated better solution quality.
Efficiency was represented by the average time the solver required to find a solution for a single satisfiable SMT file,
calculated by dividing the total wall-clock time for evaluating all found \textit{SAT} instances in the benchmark by their quantity.

We conducted all experiments involving parSAT and the SMT solvers on a Linux machine equipped with an Intel i7-12700KF processor and 32 GB of memory.
For processing each SMT file in the given benchmark, a timeout period of 600 seconds was specified.
All involved GO instances were initialized with randomly selected start values ranging between $-0.5$ and $0.5$.
We externally verified all \textit{SAT} solutions reported by parSAT by using \texttt{bitwuzla}.

\subsection{Comparison of parSAT Configurations}
In Table~\ref{tab:comp-go-algs} we present the impact of each GO method employed in parSAT to find a \textit{SAT} solution.
We executed each specified thread configuration of parSAT ten times on the Griggio benchmark and
subsequently calculated the average results.
Each row displays the percentage of the average amount of \textit{SAT} solutions found first by each employed GO algorithm
with the same number of threads used per GO method.

\begin{table}[htbp]
  \centering
  \begin{tabular}{|m{16em}|>{\raggedright\arraybackslash}m{5em}|>{\raggedright\arraybackslash}m{5em}|>{\raggedright\arraybackslash}m{5em}|}
  \hline
  & \texttt{BH} & \texttt{CRS2} & \texttt{ISRES} \\
  \hline\hline
  parsat $BH_1, CRS2_1, ISRES_1$ & 46.6\% & 27.9\% & 25.5\% \\
  \hline
  parsat $BH_2, CRS2_2, ISRES_2$ & 59.1\% & 25.2\% & 15.7\% \\
  \hline
  parsat $BH_3, CRS2_3, ISRES_3$ & 71.6\% & 17.3\% & 11.1\% \\
  \hline
  parsat $BH_4, CRS2_4, ISRES_4$ & 74.0\% & 17.3\% & 8.7\%  \\
  \hline
  parsat $BH_5, CRS2_5, ISRES_5$ & 77.5\% & 15.5\% & 7.0\% \\
  \hline
  parsat $BH_6, CRS2_6, ISRES_6$ & 83.6\% & 10.0\% & 6.4\% \\
  \hline
  parsat $BH_7, CRS2_7, ISRES_7$ & 87.0\% & 7.7\% & 5.3\% \\
  \hline
  parsat $BH_8, CRS2_8, ISRES_8$ & 87.5\% & 7.0\% & 5.5\% \\
  \hline
  parsat $BH_9, CRS2_9, ISRES_9$ & 89.6\% & 5.1\% & 5.3\% \\
  \hline
  parsat $BH_{10}, CRS2_{10}, ISRES_{10}$ & 91.2\% & 4.1\% & 4.7\% \\
  \hline
\end{tabular}
\caption{Evaluation of fastest GO algorithms per number of threads}
\label{tab:comp-go-algs}
\end{table}

The first row indicates that with one thread assigned to each GO routine,
\texttt{BH} identified more than $45\%$ of the \textit{SAT} solutions first,
while \texttt{CRS2} accounted for approximately $28\%$, and \texttt{ISRES} around $25\%$.
In the second row, with two threads per GO routine, \texttt{BH} increased its average to nearly
$60\%$, whereas \texttt{CRS2} decreased to roughly $25\%$, and \texttt{ISRES} declined to approximately $15\%$.
The subsequent rows confirm this trend: as the number of parallel GO instances increased further,
\texttt{BH} significantly improved, culminating in the final row where it was the fastest algorithm
to find a \textit{SAT} solution in over $90\%$ for the given SMT equations.
Contrarily, the share of first found \textit{SAT} solutions by \texttt{CRS2} and \texttt{ISRES} continuously diminished,
stabilizing around $4-5\%$ for 10 threads per GO instance.
This illustrates the potential of \texttt{BH} to scale and accelerate its GO process by concurrent execution which was also experienced by
Ferreiro-Ferreiro et. al~\cite{ferreiro-ferreiroBasinHoppingSynched2019}.
We hypothesize that the improved scalability of \texttt{BH} compared to \texttt{CRS2}
and \texttt{ISRES} stems from its non-population-based nature. Unlike \texttt{CRS2} and
\texttt{ISRES}, which require generating an initial population approximately 10 to 20 times
larger than the problem dimensionality and subsequently iterating over and modifying this
population during minimization, \texttt{BH} operates solely on a single initial vector whose
size matches the function's dimensionality.

Based on these findings we adjusted the number of threads assigned to each GO method.
To respect the limit of available hardware threads on the test machine, we set the maximum number of parallel running GO instances
to $20$. We used $70\%$ for running \texttt{BH} and $15\%$ each for executing \texttt{CRS2} and \texttt{ISRES},
which roughly corresponds to $14$ \texttt{BH}, $3$ \texttt{CRS2}, and $3$ \texttt{ISRES} threads.
We applied parSAT with this particular configuration ten times to the Griggio benchmark set
and measured an average runtime of $1.36$ seconds per SMT instance with approximately found
$101.8$ \textit{SAT} solutions on average. The best run of this specific parSAT configuration
discovered $102$ \textit{SAT} instances with an average runtime of $1.27$ seconds per SMT file.
This observed best result of parSAT will be used for further comparison and is denoted as $parSAT_{best} \textit{BH}_{14}, \textit{CRS2}_{3}, \textit{ISRES}_{3}$.

\subsection{Evaluation of parSAT's Performance on conventional Benchmark}
\begin{figure}[h!]
  \centering
  \includegraphics[width=0.95\textwidth]{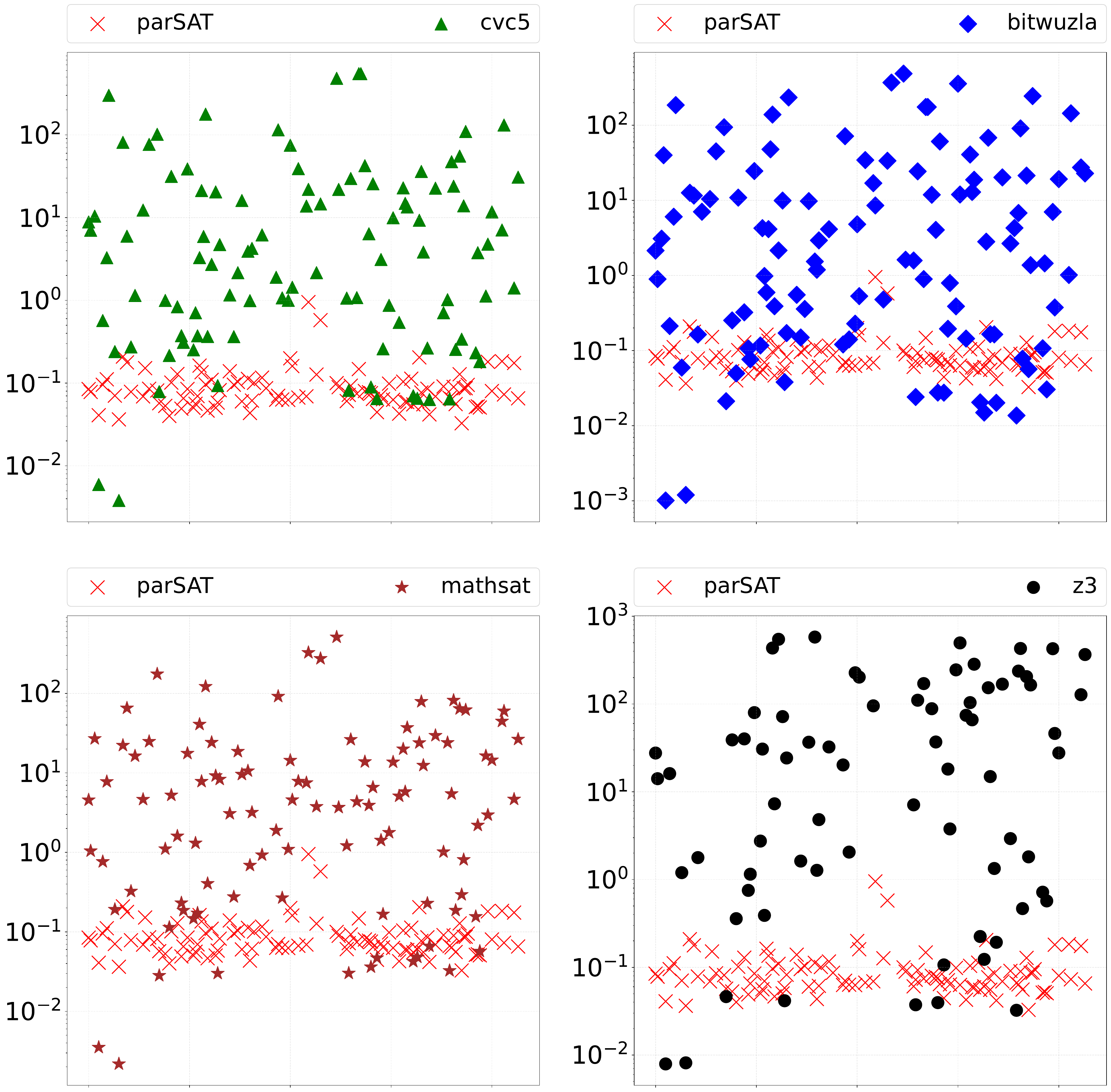}
  \caption{Comparison of solving time between parSAT and SMT solvers for satisfiable equations}
  \label{fig:parsat-smt-griggio-comparison}
\end{figure}

Figure~\ref{fig:parsat-smt-griggio-comparison} shows the runtime of the previously
described best parSAT configuration and other SMT solvers for the SMT equations
in the Griggio benchmark that were evaluated as satisfiable.
The runtime is denoted in seconds on a logarithmic scale on the Y-axis.
The X-axis represents the (virtual) index of the satisfiable SMT instance within the Griggio
benchmark.
The time parSAT took to solve each SMT instance it found \textit{SAT}
is shown by a red cross, for bitwuzla by a blue diamond, for cvc5 by a green
triangle, for MathSAT by a brown star, and for Z3 by a black dot.
Each point represents the time it took to determine an SMT equation of the
benchmark as satisfiable. If multiple points of different SMT solvers are placed exactly
on the same vertical line, this implies, that the corresponding solvers
determined the same SMT equation as \textit{SAT}.
It can be seen that parSAT did not always find a satisfiable assignment
before the compared SMT solvers, i.e., as presented on the far left side of
each plot in Figure~\ref{fig:parsat-smt-griggio-comparison}.
However, the volatility of the solving time for its identified
\textit{SAT} equations remains low while the solving time of the
other SMT solvers strongly varies.

Table~\ref{tab:comp-smt-solver-parsat-griggio} contains additional findings
on running parSAT and the SMT solvers on the Griggio benchmark.
For each solver, we present the number of found \textit{SAT} instances, \textit{UNSAT} equations,
timeouts or in case of parSAT evaluation limits, errors, and the average processing time for the found \textit{SAT} equations
in seconds.
An error indicates that the corresponding SMT solver entered an erroneous state on a given SMT
equation and terminated without issuing a \textit{SAT} or \textit{UNSAT} verdict before reaching
a timeout.
When considering the number of discovered \textit{SAT} equations,
\texttt{bitwuzla} achieved the best result by finding $108$ \textit{SAT} instances, followed
by \texttt{cvc5} that reported $104$ \textit{SAT} solutions within the given timeout.
parSAT comes at third with $102$ reported \textit{SAT} equations before \texttt{MathSAT} with
$100$, and \texttt{Z3} with $75$. Since parSAT is incomplete and cannot prove an SMT equation to be unsatisfiable,
according to \texttt{bitwuzla} or \texttt{cvc5} there are
at least $76$ guaranteed \textit{UNSAT} equations in the Griggio benchmark.
\texttt{bitwuzla} and \texttt{cvc5} ran into a timeout for $30$ instances, therefore
no final verdict can be made for the total amount of \textit{SAT} or \textit{UNSAT}
instances in the benchmark.
In terms of the average runtime per SMT file determined as \textit{SAT},
parSAT presents by far the best result with $0.1$ seconds.
The next best SMT solver is \texttt{MathSAT} with an average runtime
of $25.86$ seconds, followed by \texttt{bitwuzla} with $32.71$ seconds,
\texttt{cvc5} with $34.01$ seconds, and lastly \texttt{Z3} with $88.7$ seconds.
Overall, parSAT found approximately $6\%$ less \textit{SAT} solutions than the best solver
\texttt{bitwuzla} but required by average far less time for finding an assignment
for a satisfiable SMT equation in the Griggio benchmark.

\begin{table}[h!]
  \centering
  \begin{tabular}{|m{15em}|>{\raggedright\arraybackslash}m{2em}|>{\raggedright\arraybackslash}m{3em}|>{\raggedright\arraybackslash}m{5em}|>{\raggedright\arraybackslash}m{3em}|>{\raggedright\arraybackslash}m{5em}|}
  \hline
  & \textit{SAT} & \textit{UNSAT} & timeout / evaluation limit & errors & average \textit{SAT} runtime \\
  \hline\hline
  $parSAT_{best} BH_{14}, CRS2_{3}, ISRES_{3}$ & 102 & 0 & 112 & 0 & 0.1 \\
  \hline
  \texttt{bitwuzla} & 108 & 76 & 30 & 0 & 32.71 \\
  \hline
  \texttt{cvc5} & 104 & 76 & 30 & 4 & 34.01 \\
  \hline
  \texttt{MathSAT} & 100 & 69 & 45 & 0 & 25.86 \\
  \hline
  \texttt{Z3} & 75 & 56 & 54 & 29 & 88.7 \\
  \hline
\end{tabular}
\caption{SMT solvers and parSAT on the Griggio benchmark}
\label{tab:comp-smt-solver-parsat-griggio}
\end{table}

\subsection{parSAT's Performance on Benchmark derived from Software Verification Task}
In comparison, we show the results of running the same SMT solvers and the best
parSAT configuration on the 2019-Guedemann benchmark in
Table~\ref{tab:comp-smt-solver-parsat-guedemann}.  This benchmark was derived
from verifying an approximation of the natural logarithm and the exponential
function in a real-world scenario.  In the Cardano ledger for a proof-of-stake
based consensus protocol~\cite{kiayiasOuroborosProvablySecure2017}, precise
implementations of these two mathematical functions are required to assure that
each node of the block-chain calculates the same results.  Therefore, following
properties shall hold for an approximation of these functions

\begin{align}
    \forall x \in \mathbb{R}: & \quad x > 0 \implies  \exp'(\ln'x) \approx
  x \label{eq:prop-gm-bm1} \\
    \forall x \in \mathbb{R}: & \quad x > 0 \implies \ln'(\exp'x) \approx x \label{eq:prop-gm-bm2} \\
    \forall x, y \in \mathbb{R}: & \quad x,y \in [0, 1] \implies \exp'(x + y) \approx \exp'(x) \cdot \exp'(y) \label{eq:prop-gm-bm3} \\
    \forall x \in \mathbb{R}: & \quad x,y > 0 \implies \ln'(x \cdot y) \approx \ln'(x) \cdot
                                \ln'(y) \label{eq:prop-gm-bm4}\\
  \forall x \in \mathbb{R}: & \quad x,y \in [0,1] \implies img(x^{y}) = img(\exp'(y\cdot
                              \ln'(x))) \approx [0,1] \label{eq:prop-gm-bm5}
\end{align}

\noindent where $\exp'$ and $\ln'$ represent the approximated functions as
described in~\cite{shelleyNonIntegral}, $img$ denotes the codomain, and
$\approx$ corresponds to an absolute error of less than $\epsilon=10^{-12}$.

The implementations of these functions based on FP arithmetic together
with these properties were converted into SMT equations.
If some of these SMT equations is satisfiable, this implies that there
exists a counter-example that violates one of the stated properties.
In total 13 SMT files were generated based on these properties that constitute
the 2019-Guedemann Benchmark.

For this benchmark, both
parSAT and \texttt{bitwuzla} found $10$ \textit{SAT} instances, \texttt{cvc5}
and \texttt{MathSAT} reported $4$ and \texttt{Z3} did not find any \textit{SAT}
solution within the given time. parSAT did not find a satisfiable assignment for $3$
SMT formulas for which it reached the function evaluation limit.
\texttt{bitwuzla} denoted $2$ equations as \textit{UNSAT}
and reached for $1$ SMT equation its timeout. \texttt{cvc5} and \texttt{MathSAT}
found $1$ \textit{UNSAT} instance, and reached a timeout for $8$ SMT files.
\texttt{Z3} did not find any \textit{SAT} or \textit{UNSAT} SMT formulas, but
reached in $4$ cases the timeout.

\begin{table}[htbp]
  \centering
  \begin{tabular}{|m{15em}|>{\raggedright\arraybackslash}m{2em}|>{\raggedright\arraybackslash}m{3em}|>{\raggedright\arraybackslash}m{5em}|>{\raggedright\arraybackslash}m{3em}|>{\raggedright\arraybackslash}m{5em}|}
  \hline
  & \textit{SAT} & \textit{UNSAT} & timeout / evaluation limit & errors & average \textit{SAT} runtime \\
  \hline\hline
  $parSAT_{best} BH_{14}, CRS2_{3}, ISRES_{3}$ & 10 & 0 & 3 & 0 & 0.1 \\
  \hline
  \texttt{bitwuzla} & 10 & 2 & 1 & 0 & 37.88 \\
  \hline
  \texttt{cvc5} & 4 & 1 & 8 & 0 & 180.09 \\
  \hline
  \texttt{MathSAT} & 4 & 1 & 8 & 0 & 22.8 \\
  \hline
  \texttt{Z3} & 0 & 0 & 4 & 9 & - \\
  \hline
\end{tabular}
\caption{SMT solvers and parSAT on the 2019-Guedemann benchmark}
\label{tab:comp-smt-solver-parsat-guedemann}
\end{table}

For this benchmark, parSAT finds as many \textit{SAT} instances as \texttt{bitwuzla}
with a far lower average runtime per satisfiable SMT equation.
We see as main reason for parSAT's improved performance in the 2019-Guedemann benchmark
that the benchmark was derived from a mathematical problem where the integrated GO
algorithms in parSAT might be more efficient in finding a solution.
Another aspect could be the reduced number of variables used in the 2019-Guedemann
benchmark which vary between $1$ and $8$, whereas the Griggio benchmark contains some
SMT problems with more than $500$ variables.

\subsection{Combining parSAT with SMT Solvers}
Building on our previous findings, this section explores the potential advantages of utilizing parSAT in conjunction with an SMT solver as a new approach.
Our experimental results suggest that compared to other solvers, parSAT often
identifies satisfiable solutions faster.
Nevertheless, parSAT is unable to reliably prove an SMT formula unsatisfiable due to its inherent incompleteness.
In contrast, when no timeout occurs, SMT solvers provide \textit{SAT} or \textit{UNSAT} statements that are guaranteed to be correct,
despite potentially requiring more time than parSAT for processing.

In the subsequent theoretical consideration, we examine the potential of
combining existing SMT solvers with parSAT for handling SMT files.
Under this hypothetical framework, both the SMT solver and parSAT would concurrently initiate the search for a solution to the given SMT problem.
If parSAT identifies a satisfiable solution before the SMT solver, the solver is terminated, and a \textit{SAT} result is returned.
Conversely, if the SMT solver discovers a satisfiable assignment first, the same procedure applies to terminate parSAT.
However, if parSAT exits with \textit{UNKNOWN}, the SMT solver continues until it either proves \textit{UNSAT},
finds a satisfiable solution, or reaches a timeout. This approach appears promising as it combines the rapid execution capabilities of parSAT
with the definitive completeness properties of SMT solvers, potentially enhancing overall efficiency and reliability.

In Table~\ref{tab:com-parsat-bitwuzla}, we present the potential impact of such
a combination.  We utilized the measurements for \texttt{bitwuzla} as
representative state-of-the-art SMT solver, alongside the best observed
performance of parSAT using the
$\textit{BH}_{14}, \textit{CRS2}_{3}, \textit{ISRES}_{3}$ configuration for the
Griggio benchmark.  It is assumed that \texttt{bitwuzla} and parSAT operate
completely concurrently, without accounting for any additional overhead
introduced by parallelization.  To calculate the average solving runtime for \textit{SAT} equations,
we considered the runtime of parSAT when it returned a \textit{SAT} faster than \texttt{bitwuzla};
otherwise, we used the runtime of \texttt{bitwuzla}.
Due to parSAT's inability to reason about \textit{UNSAT} SMT equations,
instances where parSAT reached its evaluation limit
and \texttt{bitwuzla} returned \textit{UNSAT} or encountered a timeout were excluded from the
average \textit{SAT} runtime calculation.  Neither parSAT nor
\texttt{bitwuzla} encountered any errors on the Griggio benchmark.

\begin{table}[htbp]
  \centering
  \begin{tabular}{|m{16em}|>{\raggedright\arraybackslash}m{2em}|>{\raggedright\arraybackslash}m{3em}|>{\raggedright\arraybackslash}m{5em}|>{\raggedright\arraybackslash}m{3em}|>{\raggedright\arraybackslash}m{5em}|}
  \hline
  & \textit{SAT} & \textit{UNSAT} & timeout / evaluation limit & errors & average \textit{SAT} runtime \\
  \hline\hline
  $parSAT_{best} BH_{14}, CRS2_{3}, ISRES_{3}$ & 102 & 0 & 112 & 0 & 0.1 \\
  \hline
  \texttt{bitwuzla} & 108 & 76 & 30 & 0 & 32.71 \\
  \hline
  $parSAT_{best}$ + \texttt{bitwuzla} & 113 & 76 & 25 & 0 & 12.75 \\
  \hline
\end{tabular}
\caption{Combination of parSAT with \texttt{bitwuzla} on the Griggio benchmark}
\label{tab:com-parsat-bitwuzla}
\end{table}

The table compares the outcomes of executing parSAT and \texttt{bitwuzla} both
independently and in a combined approach.  It provides data on
identified \textit{SAT} instances, \textit{UNSAT} formulas, and the average
runtime per found \textit{SAT} SMT file, measured in seconds. The first two rows display the
results of running parSAT and \texttt{bitwuzla} separately, while the third row
illustrates the potential effect of systematically combining the two methods.
Based on the recorded data, and in contrast to their independent application,
the combined approach would increase the number of identified \textit{SAT}
instances to $113$ and reduce the timeouts to $25$.  Moreover, the
average solving runtime for satisfiable equations would decrease to $12.75$ seconds,
representing a reduction of slightly more than $60\%$ compared to exclusively using \texttt{bitwuzla}.

Our research suggests a promising avenue for advancing this approach by
leveraging the evaluated data sets by parSAT after it reached its
function evaluation limit. The SMT solver could then utilize these preprocessed
values to enhance its internal lemma generation capabilities and exclude already
unsatisfiable assignments from its potential solution set, thereby accelerating
the overall solving process.

%%% Local Variables:
%%% mode: latex
%%% TeX-master: "main"
%%% End:

%% file: 06_conclusion.tex
\section{Conclusion and Outlook}
\label{sec:conclusion}

Experiments with parSAT demonstrate that its portfolio-approach utilizing
multiple different GO algorithms competing concurrently to find solutions,
presents a considerable alternative to current state-of-the-art SMT solvers
to quickly verify if a given SMT formula might be satisfiable.
Furthermore, parSAT successfully handled and
processed a variety of real-world SMT equations with either small or large
numbers of variables. The modular architecture of parSAT enables straightforward
integration of additional GO routines that could further enhance the
complementary and competitive solving approach. In particular, we demonstrated
parSAT's ability to find \textit{SAT} solutions to some SMT instances derived from
practical problems with mathematical nature faster than current state-of-the-art SMT solvers.

Because of the currently integrated GO methods,
parSAT is a semi-decision procedure that underapproximates the search space. If
it finds a solution satisfying a problem instance, it is guaranteed to be
correct but might label an actual satisfiable formula as \textit{UNKNOWN}. In the
current version, due to some implementation details of the integrated GO approaches,
parSAT is limited on finding solutions consisting of finite FP values
only. Even though parSAT's GO routines could potentially calculate $\pm \infty$ values for
variables in single precision format, they currently cannot systematically find
\textit{NaN}-values in general or $\pm\infty$ in double precision.

We plan to integrate more types of GO algorithms into parSAT, such as
deterministic variants or brute-force based approaches. We also see a potential
benefit in implementing a specific GO method that evaluates the optimization
function by allowing the controlled selection of non-finite FP numbers as
input. Additionally, we aim to further investigate how the generation of the
optimization function could be improved such that GO algorithms might find the
global minimum faster and in a more reliable way.
Another aspect to consider is the potential acceleration of integrated GO methods
through further parallelization, which may involve utilizing CPU-specific vector instructions,
GPU-specific implementations, or deploying GO instances to a distributed cluster
with scalable computing capabilities based on cloud infrastructure.
The approach itself, formulating a constraint problem as a GO
function, is not limited to FP only; we aim to support more theories defined by
the \smtlib{} standard, such as bitvector or integer theory.

In addition, we also plan to extend the use of the approach in two domains. The
first is the evaluation how parSAT might be used in a complementary approach
with existing SMT solvers. To facilitate this process, we will investigate how
equalities can be derived to propagate information to other theory solvers
during theory combination. The second promising topic is the application of the
techniques used in parSAT to the problem domain of SMT model
counting. This involves counting the number of all satisfiable assignments for a
given SMT equation.